\documentclass[twocolumn]{aastex7}
\usepackage{amssymb,amsmath,amsthm}

\shorttitle{Nearest Billion-Solar-Mass Black Hole}
\shortauthors{Wrobel, Pesce \& Nyland}

\begin{document}

\title{Inside the Stagnation Radius of the Nearest 
Billion-Solar-Mass Black Hole}

\author[0000-0001-9720-7398]{J. M. Wrobel}
\affiliation{National Radio Astronomy Observatory, P.O. Box O,
  Socorro, NM 87801, USA}
\email[show]{jwrobel@nrao.edu}

\author[0000-0002-5278-9221]{D. W. Pesce}
\affiliation{Center for Astrophysics $|$ Harvard \& Smithsonian, 60
Garden Street, Cambridge, MA 02138, USA}
\affiliation{Black Hole Initiative, Harvard University, 20 Garden
Street, Cambridge, MA 02138, USA}
\email{dpesce@cfa.harvard.edu}

\author[0000-0003-1991-370X]{K. E. Nyland}
\affiliation{U.S. Naval Research Laboratory, 4555 Overlook Ave SW, 
  Washington, DC 20375, USA}
\email{kristina.nyland.civ@us.navy.mil}

\correspondingauthor{J. M. Wrobel}

\received{2025 May 16}\accepted{2025 August 9}\submitjournal{ApJ}

\begin{abstract}
We used the NSF Jansky Very Large Array at a frequency $\nu =$ 22\,GHz
to study the nearest billion-solar-mass black hole, in the early-type
galaxy NGC\,3115 at a distance of 9.7\,Mpc. We localize a faint
continuum nucleus, with flux density $S_{\rm 22\,GHz} =
48.2\pm6.4\,\mu$Jy, to a FWHM diameter $d_{\rm 22\,GHz} <$ 59\,mas
(2.8\,pc). We find no evidence for adjacent emission within a
stagnation region of radius $R_{\rm sta} \sim$ 360\,mas (17\,pc)
identified in a recent hydrodynamic simulation tailored to NGC\,3115.
Within that region, the simulated gas flow developed into an
advection-dominated accretion flow (ADAF). The nucleus' luminosity
density $L_{\rm 22\,GHz} = 5.4 \times 10^{17}\,\rm W\,Hz^{-1}$ is
about 60 times that of Sagittarius\,A$^\star$. The nucleus' spectral
index $\alpha_{\rm 10\,GHz}^{\rm 22\,GHz} = -1.85\pm0.18$ ($S_\nu
\propto \nu^\alpha$) indicates optically-thin synchrotron
emission. The spectral energy distribution of the nucleus peaks near
$\nu_{\rm peak} =$ 9\,GHz. Modeling this radio peak as an ADAF implies
a black hole mass $M_{\rm ADAF} = (1.2\pm0.2) \times 10^9\,M_\odot$,
consistent with previous estimates of $(1-2) \times 10^9\,M_\odot$
from stellar or hot-gas dynamics. Also, the Eddington-scaled accretion
rate for NGC\,3115, $\dot{M}_{\rm ADAF}/\dot{M}_{\rm Edd} =
1.2^{+1.0}_{-0.6} \times 10^{-8}$, is about 4-8 times lower than
recent estimates for Sagittarius\,A$^\star$.
\end{abstract}

\keywords{Accretion (14); Active galactic nuclei (16); Early-type
galaxies(429); Supermassive black holes (1663); Interferometry (808)}

\section{Motivation}

Supermassive black holes (BHs) spend the majority of their time
accreting at well below the Eddington rate $\dot{M}_{\rm Edd} = 2.2
\times 10^{-8}\,M\,M_\odot\,yr^{-1}$, where $M$ is the BH mass in Solar
units \citep[e.g.,][]{hec14}. In the local Universe this accretion
state manifests as low-luminosity active galactic nuclei
\citep[LLAGNs; e.g.,][]{ho08,ho09}. For LLAGNs, it is thought that the
material near the BHs follows the advection-dominated accretion flow
(ADAF) solution to the hydrodynamic equations for viscous and
differentially rotating flows \citep[e.g.,][]{yua14}. Direct imaging
of such accretion flows is beyond current observational capabilities 
and strongly motivates the design of future facilities, such as the
next-generation Event Horizon Telescope \citep[ngEHT;][]
{joh23,doe23,nai24}.

Further from the BHs, the ADAFs may be fed by Bondi accretion from the
hot atmospheres of their host spheroids \citep[e.g.,][]{yua14}.  In a
few cases X-ray facilities can spatially resolve the Bondi regions
\citep[e.g.,][]{ina20}, informing theoretical studies of how the gas
densities ($\rho$), temperatures, and accretion rates ($\dot{M}$) vary
with radius $R$ from the BH \citep[e.g.,][]{ina18,cho23,guo23}. For
example, a measured $\rho(R)$ can be compared with $\rho(R) \propto
R^{-1.5}$ from Bondi theory. Profiles flatter than Bondi may suggest
that the accretion flows are directing some of their mass away from
the BHs and suppressing their ability to radiate -- the hallmark of
LLAGNs.

Such generic comparisons can provide guidance about accretion flow
suppression. But stronger insights will follow from realistic
calculations or simulations that are tailored to the unique settings
of individual LLAGNs, such as accretion onto Sagittarius\,A$^\star$
via the stellar winds of nearby Wolf-Rayet stars
\citep[e.g.,][]{res18}.

Here, we focus on NGC\,3115 at a distance $D =$ 9.7\,Mpc \citep{ton01}
as it hosts the nearest billion-solar-mass BH, with a fiducial mass
$M_{\rm fid} \sim 1.5 \times 10^9\,M_\odot$ from stellar or hot-gas
dynamics \citep{kor96,ems99,won11,won14}. The LLAGN of this 
early-type galaxy (ETG) was discovered with the NSF Very Large Array
\citep[VLA;][]{tho80} in its A configuration at a frequency $\nu =$
8.5\,GHz \citep{wro12}. It has a flux density $S_{\rm 8.5\,GHz} =
290\pm30\,\mu$Jy (corresponding to a luminosity density $L_{\rm
8.5\,GHz} = 3.3 \times 10^{18}\,\rm W\,Hz^{-1}$ or $L_{\rm 8.5\,GHz} 
= 3.3 \times 10^{25}$\,erg\,s$^{-1}$\,Hz$^{-1}$) and a diameter
$d_{\rm 8.5\,GHz} <$ 170\,mas (8.0\,pc). For context at 8.5\,GHz,
\citet{kra02} and \citet{cap09} used the VLA in its A configuration to
search for LLAGN in dozens of optically selected ETGs, but their
detection thresholds precluded discovery of analogs of the LLAGN in
NGC\,3115.

Confusion from NGC\,3115’s population of X-ray binaries means that the
LLAGN cannot be assigned a counterpart at 2-10\,keV
\citep{wro12,won14}. This implies an X-ray-to-Eddington ratio of less
than $2.1 \times 10^{-9}$ assuming a radiative efficiency $\eta = 0.1$
for the Eddington luminosity. Applying the \citet{ho09} bolometric
correction leads to a bolometric-to-Eddington ratio $L_{\rm
  bol}/L_{\rm Edd} < 3.3 \times 10^{-8}$, an extreme ratio rare among
LLAGNs in optically selected galaxies. NGC\,3115's LLAGN also lacks
evidence for optical emission lines like H$\alpha$, H$\beta$, and
\ion{O}{3} \citep{ho03,gue16}. Such galaxies are referred to as having
absorption-line or passive nuclei \citep{ho03,nyl16}.

The fiducial $M_{\rm fid}$ and the hot-atmosphere sound speed 
\citep{won11,won14} for NGC\,3115 imply a Bondi radius $R_{\rm Bon} =
3\farcs6$ (170\,pc). \citet{yao20} performed 2D hydrodynamic
simulations of gas flows tailored to NGC\,3115's well-studied traits
within $5\,R_{\rm Bon}$. They identified a stagnation region of radius
$R_{\rm sta} \sim 0.1\,R_{\rm Bon} \sim$ 360\,mas (17\,pc): the
dominant gas motions were inflows inside $R_{\rm sta}$ and outflows
beyond $R_{\rm sta}$. 1D calculations also tailored to NGC\,3115
identified a similar stagnation region \citep{shc14}. Inside $R_{\rm
sta}$ the \citet{yao20} simulation developed two ADAF traits, namely
gas densities that varied with radius $R$ as $\rho(R) \propto
R^{-0.8}$ and a geometrically-thick disk containing hot electrons and
hotter ions. The accretion also mainly occurred within the polar
regions, a trait also recognized in a tailored 3D hydrodynamic
simulation of Sagittarius\,A$^\star$ \citep{res18}.

\citet{yao20} did not consider magnetic fields, so the emissive
properties of their simulated ADAF cannot be compared to the LLAGN
\citep{wro12}. But that the simulation developed an ADAF does motivate
another approach to studying NGC\,3115: assemble the LLAGN's spectral
energy distribution (SED) and model it using ADAF theory
\citep[e.g.,][]{NY95a,NY95b,mah97,nem14,ban19,pes21}.  It might be
that both the outflow-dominated and inflow-dominated regions are
emitting. Our focus is on the SED of the latter. To mitigate
contamination from emission beyond $R_{\rm sta}$, the SED should
ideally be assembled at high resolutions, taken to mean resolutions at
full width half maximum (FWHM) of less than $2R_{\rm sta} \sim$
720\,mas (34\,pc).

\citet{alm18} pioneered the SED approach for NGC\,3115, but had only
one radio detection of the LLAGN at a high resolution \citep{wro12}.
The \citet{yao20} simulation guides us to seek more detections of the
LLAGN at high resolution ($< 2R_{\rm sta}$) and to treat data at low
resolution ($> 2R_{\rm sta}$) as upper limits on the LLAGN. In Section
\ref{sec:data} of this paper, we report a new, high-resolution radio
detection of the LLAGN in NGC\,3115. In Section \ref{sec:inside} we
use the new and literature data to better define the properties
of the LLAGN inside $R_{\rm sta}$. We also briefly touch on conditions
beyond $R_{\rm sta}$ in Section \ref{sec:beyond}. We close, in Section
\ref{sec:summary}, with a summary and conclusions.

Throughout, uncertainties are reported as $1\sigma$ unless stated
otherwise and literature distances are converted to $D =$ 9.7\,Mpc
\citep{ton01} when necessary.

\section{Data}\label{sec:data}

NGC\,3115 was observed on 2024 December 19 and 20 UT with the A
configuration of the NSF Karl G.\ Jansky Very Large Array 
\citep[JVLA;][]{per11}. A coordinate equinox of 2000 was employed.
J1007$-$0207, at a position $\alpha(J2000) = 10^{h} 07^{m} 04\fs34992$,
$\delta(J2000) = -02\arcdeg 07\arcmin 10\farcs9177$ and with a 1D 
position error of 2.0\,mas, was used as a gain calibrator. The
switching time between it and NGC\,3115 was 4\,m and involved a
switching angle of 5.6\arcdeg. Reference pointing was implemented.

\begin{figure*}[tbh]
\plotone{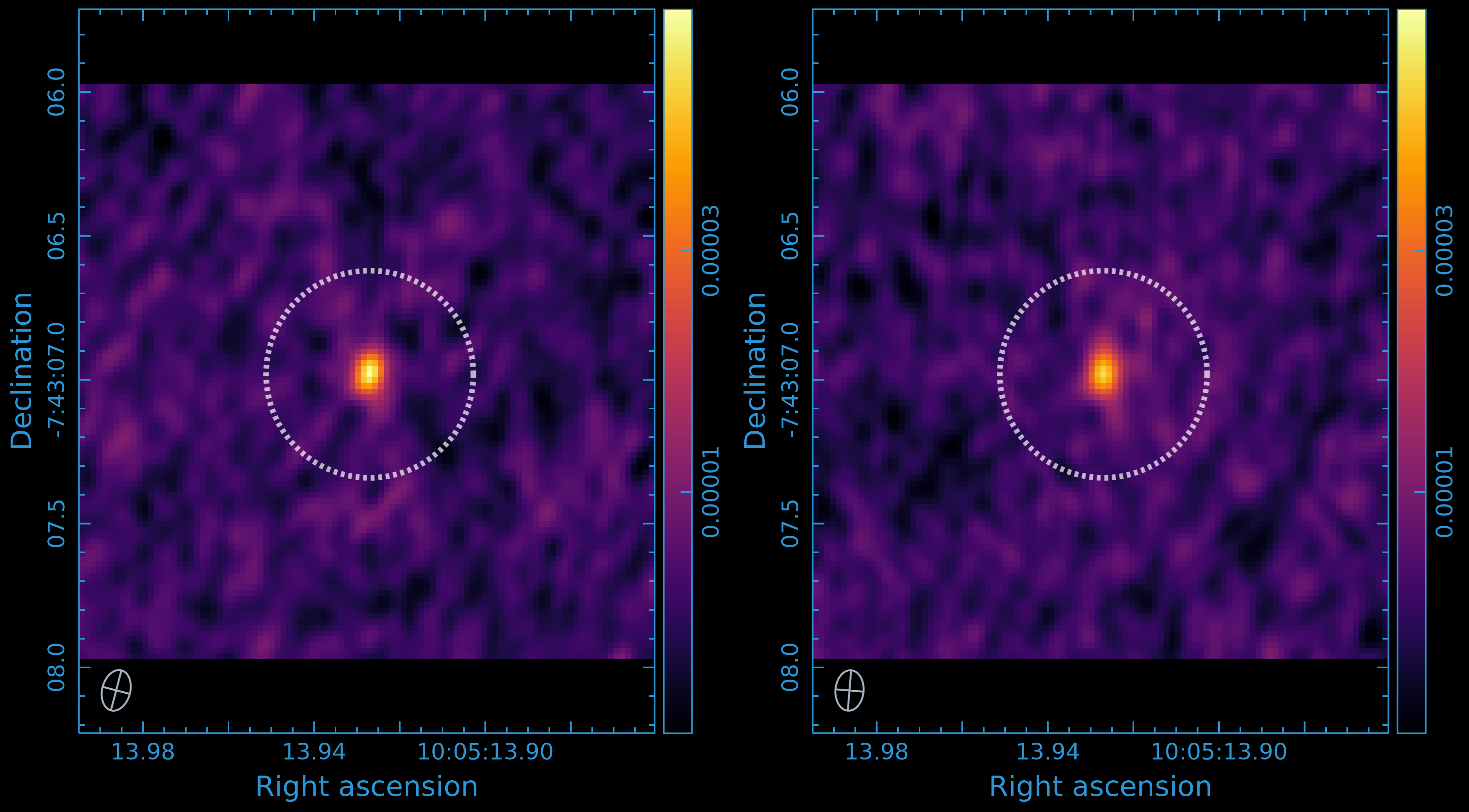}
\caption{JVLA images of the Stokes $I\/$ emission at $\nu =$ 22\,GHz
  from NGC\,3115, visualized via The Cube Analysis and Rendering Tool
  for Astronomy \citep[CARTA;][]{com21}. The scale is $1\arcsec =$
  47\,pc.  The ellipses show the synthesized beam dimensions at FWHM,
  characterized by their major axes $\theta_{\rm maj}$, minor axes
  $\theta_{\rm min}$, and elongation position angles. The ellipses
  share the same geometric angular resolution $\theta^{\rm
    22\,GHz}_{\rm geo} = \sqrt{\theta_{\rm maj} \times \theta_{\rm
      min}} =$ 120\,mas (5.6\,pc).  The dashed circle shows the
  stagnation region diameter $2R_{\rm sta}$ from the \citet{yao20}
  simulation. The color bar linearly spans $-10$ to $+50\,\mu\rm
  Jy\,beam^{-1}$. Left: UT date is 2024 December 19.  Root-mean-square
  (RMS) noise is $3.0\,\mu\rm Jy\,beam^{-1}$. Right: UT date is 2024
  December 20. RMS noise is $3.2\,\mu\rm Jy\,beam^{-1}$.}
\label{fig:images}
\end{figure*}

Data were acquired in dual circular polarizations, each spanning 4 
$\times$ 2\,GHz of bandwidth centered at $\nu =$ 22\,GHz. 3C\,286 was
observed to set the amplitude scale to an estimated accuracy of 10\%.
This estimate stems from the modest elevation differences of
9-16\arcdeg\, between observations of 3C\,286 and the gain calibrator,
and conservatively doubles the base level suggested by \citet{per17}.
The exposure time on NGC\,3115 was about 1.5\,h per observation.
Polarization calibration was not implemented, as each day's
observation spanned only a few hours.

The visibility data were pipeline calibrated and custom imaged using
versions 6.6.1 and 6.6.4-34, respectively, of the Common Astronomy 
Software Applications package \citep{cas22}. For each observation, task
{\tt tclean} was used to form an image of the Stokes $I\/$ emission
from NGC\,3115 by weighting the visibility data with a Briggs
robustness of 0.5 and invoking the mtfs deconvolver with a 
straight-line spectral model (Figure \ref{fig:images}). The images
spanned 100 pixels per axis and had 20\,mas pixels. Each image
independently detects the nucleus at $\nu =$ 22\,GHz and locates its
peak at a position $\alpha(J2000) = 10^{h} 05^{m} 13\fs927$, 
$\delta(J2000) = -07\arcdeg 43\arcmin 06\farcs98$. The position error
is dominated by phase-referencing strategies and a 1D estimate for it
is about 100\,mas.

\begin{figure}[tbh]
\plotone{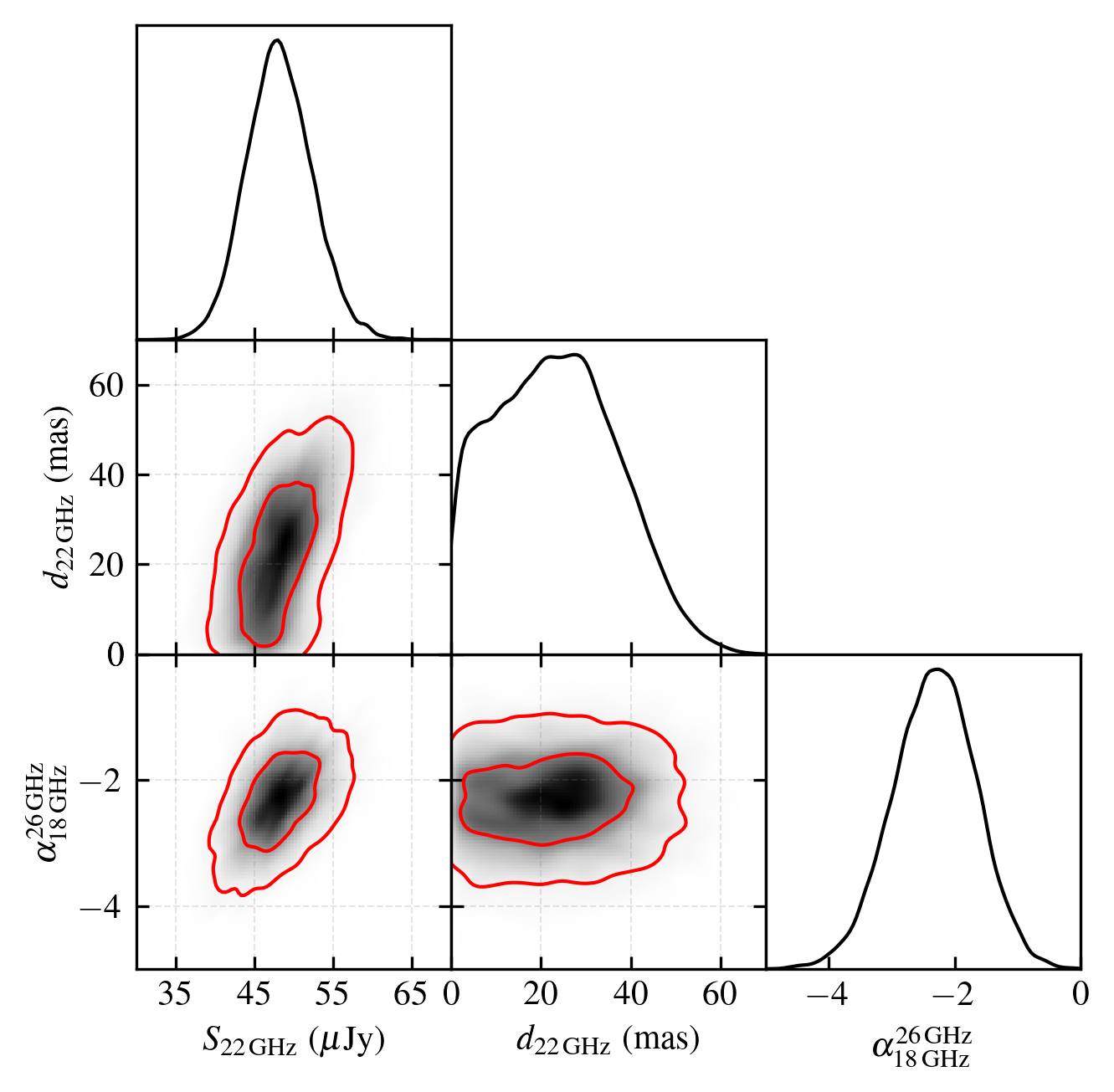}
\caption{Posterior distribution of a circular Gaussian's flux density 
$S_{\rm 22\,GHz}$, FWHM diameter
$d_{\rm 22\,GHz}$, and a linear model for 
the in-band spectral index $\alpha_{\rm 18\,GHz}^{\rm 26\,GHz}$. Inner
and outer contours enclose probabilities of 50\% and 90\%, 
respectively.}
\label{fig:Setc}
\end{figure}

We combined the calibrated visibility data from both observations at
$\nu =$ 22\,GHz and jointly fit models to them, assuming no changes in
the source between the two observations (Figure \ref{fig:Setc}). (The
source size modeling procedure is detailed in
\autoref{app:SourceSizeModeling}.) A Gaussian source model for the
nucleus' flux density found it to be (1) faint, with $S_{\rm 22\,GHz}
= 48.2\pm6.4\,\mu$Jy, where the error is the quadratic sum of the
modeling uncertainty and the 10\% amplitude-scale uncertainty; and (2)
point-like, with a FWHM diameter $d_{\rm 22\,GHz} <$ 59\,mas (99.7\%
confidence region). A linear model for the nucleus' in-band spectral
index found $\alpha_{\rm 18\,GHz}^{\rm 26\,GHz} = -2.3\pm0.7$ ($S_\nu
\propto\nu^\alpha$), broadly indicating a synchrotron origin.  The
large error in $\alpha_{\rm 18\,GHz}^{\rm 26\,GHz}$ is mainly due to
the source’s limited signal-to-noise ratio. The above modeling
results, rather than the images in Figure \ref{fig:images}, will be
used hereafter to characterize the nucleus.

With adequate signal-to-noise, structures as large as $1\farcs2$ could
be represented at $\nu =$ 22\,GHz. But no such structures were found 
after tapering the visibility data to achieve 
$\theta^{\rm 22\,GHz}_{\rm geo} =$ 300\,mas and 520\,mas, with 
off-nuclear RMS noise levels of $4.9\,\mu\rm Jy\,beam^{-1}$ and 
$6.9\,\mu\rm Jy\,beam^{-1}$, respectively. Also, the nuclear flux
densities from the tapered images agreed with the $S_{\rm 22\,GHz}$
inferred from the Gaussian source model.

\begin{figure}[tbh]
\plotone{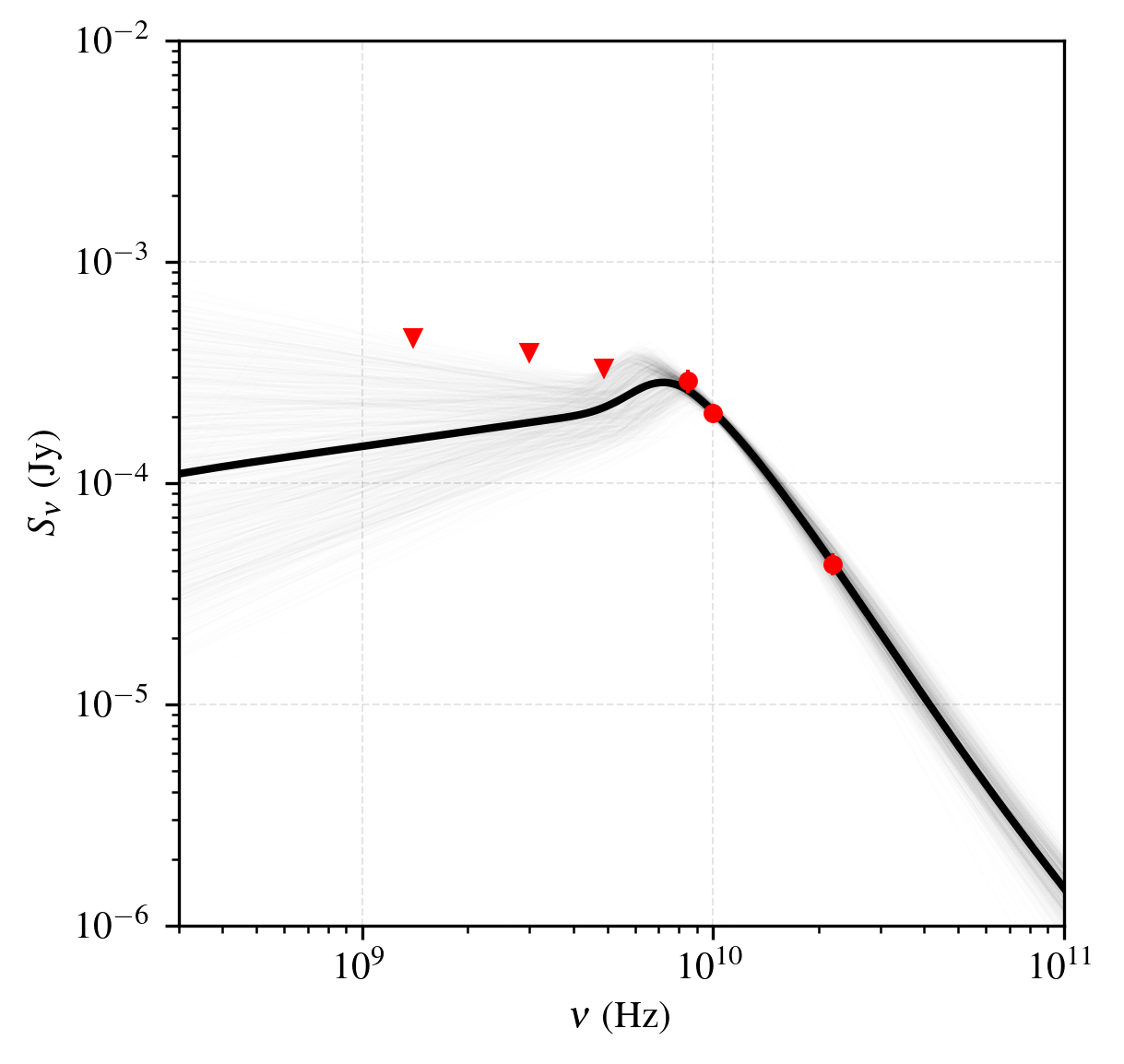}
\caption{Broadband spectrum of the LLAGN in NGC\,3115. Triangles
indicate 3$\sigma$ upper limits with $\theta^\nu_{\rm geo} > 2R_{\rm
sta}$. Circles with 1$\sigma$ error bars indicate detections with 
$\theta^\nu_{\rm geo} < 2R_{\rm sta}$.  The faint
gray lines are individual samples from the posterior distribution, 
which illustrate the full range of model behavior. The thick black 
line shows the posterior average, which is a fit to the thermal 
ADAF model presented in Section~3.3.}
\label{fig:broadband}
\end{figure}

Table \ref{tab:augment} augments the new $S_{\rm 22\,GHz}$ with
previous continuum photometry for NGC\,3115 at lower radio
frequencies, including 3$\sigma$ upper limits at low resolution
$\theta^\nu_{\rm geo} > 2R_{\rm sta}$ and detections at high
resolution $\theta^\nu_{\rm geo} < 2R_{\rm sta}$. Figure
\ref{fig:broadband} shows the resulting broadband spectrum. The
nucleus has a band-to-band spectral index $\alpha_{\rm 10\,GHz}^{\rm
  22\,GHz} = -1.85\pm0.18$, indicating optically-thin synchrotron
emission and reinforcing the synchrotron origin suggested from the
in-band's $\alpha_{\rm 18\,GHz}^{\rm 26\,GHz} = -2.3\pm0.7$ (Figure
\ref{fig:Setc}).

\begin{deluxetable*}{lccccccc}\label{tab:augment}
\tablecolumns{8} \tablewidth{0pc} 
\tablecaption{Photometry for NGC\,3115's Broadband Spectrum} 
\tablehead{
  \colhead{UT}           & \colhead{Frequency}    &
  \colhead{Resolution}   & \colhead{Resolution}   &
  \colhead{Local RMS}    & \colhead{Flux Density} &
  \colhead{Luminosity Density} & \colhead{Ref.}   \\
  \colhead{Date} & \colhead{$\nu$ (GHz)}  &
  \colhead{$\theta^\nu_{\rm geo}$ (mas)}  & 
  \colhead{$\theta^\nu_{\rm geo}$ (pc)}   &
  \colhead{$\mu$Jy\,beam$^{-1}$}          &
  \colhead{$S_\nu$ ($\mu$Jy)}             & 
  \colhead{$L_\nu$ (W\,Hz$^{-1}$)} & \colhead{} \\ 
  \colhead{(1)} & \colhead{(2)} & \colhead{(3)} &
  \colhead{(4)} & \colhead{(5)} & \colhead{(6)} &
  \colhead{(7)}}
\startdata
2002 Aug 12& 1.4& 5900& 280& 150& $<$ 450& $< 5.1\times10^{18}$& 1\\
2021 Dec 14& 3.0& 2800& 130& 130& $<$ 390& $< 4.4\times10^{18}$& 2 \\
1987 Feb 1 & 4.9& 8500& 400& 110& $<$ 330& $< 3.7\times10^{18}$& 3 \\
2004 Nov 16& 8.5&  300&  14&  18& 290$\pm$30& $3.3\times10^{18}$& 4 \\
2015 Jun 12& 10 &  240&  11& 4.4& 207$\pm$10& $2.3\times10^{18}$& 5 \\
2024 Dec 19-20&22&120&5.6& 3.1& 48.2$\pm$6.4& 5.4$\times10^{17}$& 6 \\ 
\enddata
\tablecomments{References. (1) \citet{whi97}; (2) \citet{lac20};
(3) \citet{fab89}; (4) \citet{wro12}; (5) \citet{jon19}; (6) this
work.}
\end{deluxetable*}

\section{Inside the Stagnation Radius}\label{sec:inside}

Adopting a distance $D =$ 9.7\,Mpc for NGC\,3115 \citep{ton01}, 
1\arcsec\, subtends 47\,pc. Gas motions are predominantly inflows
inside the stagnation radius $R_{\rm sta} \sim$ 360\,mas (17\,pc)
identified in the simulation tailored to NGC\,3115 \citep{yao20}.

\subsection{Physical Traits}\label{traits}

NGC\,3115's point-like continuum nucleus -- its LLAGN -- is localized
to a FWHM diameter  $d_{\rm 22\,GHz} <$ 59\,mas 
(2.8\,pc). This is three times smaller than previous localizations 
based on images near 9\,GHz \citep{wro12,jon19}. Assuming that the 
LLAGN at $\nu =$ 22\,GHz coincides with the BH, the localization 
radius corresponds to less than $10^4$ times the Schwarzschild radius
$R_{\rm Sch} = 2GM_{\rm fid}/c^2$ of the fiducial BH mass $M_{\rm fid}
\sim 1.5 \times 10^9\,M_\odot$. The FWHM diameter and flux density of
the LLAGN imply a brightness temperature higher than 240\,K at $\nu =$
22\,GHz. 

The position of the LLAGN at $\nu =$ 22\,GHz agrees to within 50\,mas
with those reported near $\nu =$ 9\,GHz \citep{wro12,jon19}, which in
turn coincide with the position of the galaxy's photocenter measured
in the infrared K-band \citep{skr06,jon19}. In the optical V-band the
stellar nucleus of NGC\,3115 is spatially resolved and has a half-light
diameter $d_{\rm hal} =$ 108\,mas \citep[5.1\,pc;][]{kor96,ems99}. The
1D calculations and the 2D hydrodynamic simulations tailored to
NGC\,3115 assume that the BH is centered on the stellar nucleus
\citep{shc14,yao20}. But the positional accuracy of the stellar
nucleus \citep[$\pm$ 500\,mas at 95\% confidence;][]{nor14} is
insufficient to say if it coincides with the LLAGN.

The LLAGN appears to be isolated at $\nu =$ 22\,GHz, with no evidence
for off-nuclear emission on scales from 120\,mas (5.6\,pc) to 
$1\farcs2$ (56\,pc). This range fully samples the diameter $2R_{\rm
st} \sim$ 720\,mas (34\,pc) of the stagnation region identified in the
\citet{yao20} simulation, where the dominant gas motions are inflows.
Isolation of the LLAGN was also noted from moderately deep images near
$\nu =$ 9\,GHz with high resolution, $\theta^\nu_{\rm geo} < 2R_{\rm
sta}$ (Table \ref{tab:augment}).

NGC\,3115's LLAGN has a steep spectrum indicative of synchrotron
emission, measured in-band $\alpha_{\rm 18\,GHz}^{\rm 26\,GHz} =
-2.3\pm0.7$ and band-to-band $\alpha_{\rm 10\,GHz}^{\rm 22\,GHz} =
-1.85\pm0.18$.  The latter statement assumes flux density stability
between 9.5\,yr, a trait previously found to hold near $\nu =$ 9\,GHz
between 10.6\,yr \citep{jon19}. Still, if the band-to-band index is
compromised by time variability, then much of the radiating material
would need to occupy a volume, set by light travel times, of diameter
$\lesssim$ 19\,light yr (5.8\,pc). Such an upper limit would be
consistent with the measured diameter $d_{\rm 22\,GHz} <$ 59\,mas
(2.8\,pc) and further underscore the compactness of the LLAGN.

At $\nu =$ 22\,GHz, NGC\,3115's synchrotron nucleus has a luminosity
density $L_{\rm 22\,GHz} = 5.4 \times 10^{17}\,\rm W\,Hz^{-1}$. This
is only about 60 times that of Sagittarius\,A$^\star$, which has a
distance $D =$ 8.1\,kpc \citep{rei19,gra19} and a non-flaring flux
density $S_{\rm 21.2\,GHz} =$
1.16\,Jy\footnote{https://science.nrao.edu/science/service-observing}.

\subsection{Context from Bright Galaxy Surveys}\label{context}

Continuum surveys of optically bright galaxies like NGC\,3115 have not
been made with the VLA or JVLA in its A configuration at $\nu =$
22\,GHz. However, at a nearby frequency, 15\,GHz, the Palomar sample of
galaxies with optical emission-line nuclei \citep{ho03} was surveyed
in the A configuration with $\theta^{\rm 15\,GHz}_{\rm geo} =$
150\,mas \citep{nag05,sai18}. Galaxy distances extended up to 120\,Mpc
so the linear resolutions were 87\,pc or finer. NGC\,3115 is in the 
defining Palomar sample but was not targeted at $\nu =$ 15\,GHz
because it lacks optical emission lines \citep{ho03,gue16}.
NGC\,3115's $S_{\rm 22\,GHz}$ and $\alpha_{\rm 10\,GHz}^{\rm 22\,GHz}$
imply $S_{\rm 15\,GHz} \sim 100\,\mu$Jy, leading to a luminosity
density $L_{\rm 15\,GHz} \sim 1.1 \times 10^{18}\,\rm W\,Hz^{-1}$.
This is about a factor of two below the least-luminous of the 112
emission-line nuclei detected in the $\nu =$ 15\,GHz survey
\citep{nag05,sai18}. 

Comparing NGC\,3115 with other absorption-line nuclei may offer more
insights. At $\nu =$ 1.5\,GHz, a Northern subset of the defining
Palomar sample \citep{ho03} was surveyed with e-MERLIN with 
$\theta^{\rm 1.5\,GHz}_{\rm geo} =$ 200\,mas \citep{bal18,bal21a}.
Again, galaxy distances extended up to 120\,Mpc so the linear
resolutions were 120\,pc or finer. Of the 28 absorption-line nuclei
targeted, 5 have point-like detections with a typical luminosity
density $L_{\rm 1.5\,GHz} \sim 10^{20}\,\rm W\,Hz^{-1}$ and 23 have
luminosity densities $L_{\rm 1.5\,GHz} < (0.6-60) \times 
10^{18}\,\rm W\,Hz^{-1}$. The upper limit on NGC\,3115's luminosity
density at $\nu =$ 1.4\,GHz (Table \ref{tab:augment}) thus seems
typical for an absorption-line nucleus.

At $\nu =$ 5\,GHz, a Northern subset of the ATLAS$^{\rm 3D}$ sample of
ETGs \citep{cap11} was surveyed with the JVLA in its A configuration
with $\theta^{\rm 5\,GHz}_{\rm geo} =$ 500\,mas \citep{nyl16}. Galaxy
distances extended up to 42\,Mpc so the linear resolutions were
100\,pc or finer. The rates of point-like detections are higher among
emission-line nuclei (70/101 $\sim$ 69\%) than among absorption-line
nuclei (6/47 $\sim$ 13\%), again underscoring the importance of
comparing NGC\,3115 with other absorption-line nuclei.  For the 41
undetected absorption-line nuclei in Atlas$^{\rm 3D}$ galaxies, the
typical constraints on their luminosity densities are $L_{\rm 5\,GHz}
< (1-10) \times 10^{18}\,\rm W\,Hz^{-1}$.  Thus the upper limit on
NGC\,3115's luminosity density at 4.9\,GHz (Table \ref{tab:augment})
appears to be typical for an absorption-line nucleus.

Using the aforementioned facilities at $\nu =$ 1.5-50\,GHz, it would
be straightforward to constrain the synchrotron spectrum of NGC\,3115
with high resolutions, that is, with $\theta^\nu_{\rm geo} < 2R_{\rm
  sta} \sim$ 720\,mas (34\,pc). The most pressing need is for
high-resolution data below 10 GHz.

\subsection{ADAF Model}\label{adaf}

\begin{figure}[tbh]
\plotone{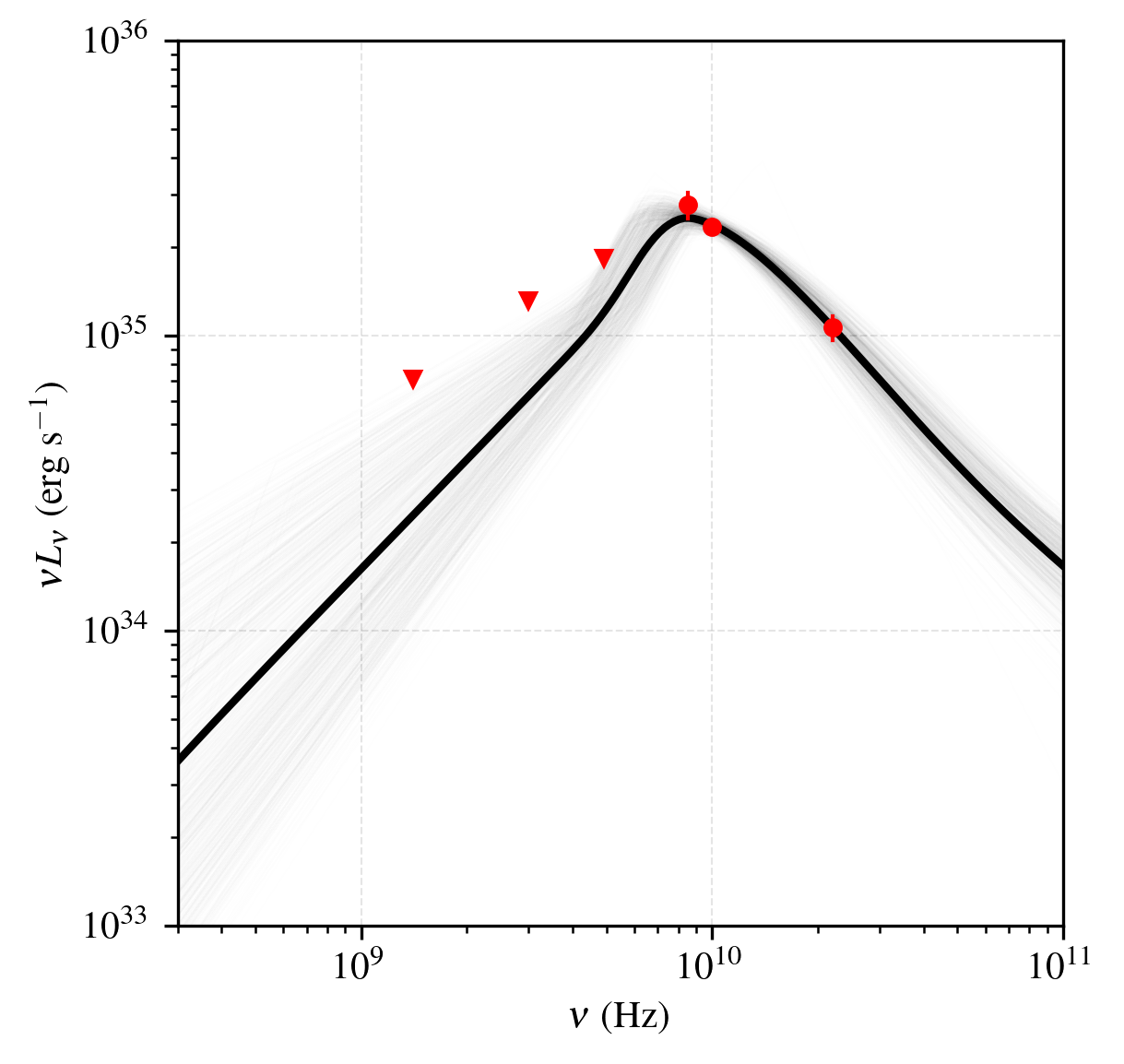}
\caption{$\nu L_\nu$ SED of the LLAGN in NGC\,3115, converted from
  Figure \ref{fig:broadband} using a distance $D =$ 9.7\,Mpc.  The
  faint gray lines are individual samples from the posterior
  distribution, which illustrate the full range of model behavior. The
  thick black line shows the posterior average, which is a fit to the
  thermal ADAF model presented in \autoref{adaf}.}
\label{fig:sed}
\end{figure}

We converted NGC\,3115's broadband spectrum in Figure
\ref{fig:broadband} to the $\nu L_\nu$ SED in Figure \ref{fig:sed}.
The SED of the LLAGN has an obvious synchrotron peak near $\nu_{\rm
  peak} =$ 9\,GHz. The LLAGN has had its extent constrained to a FWHM
diameter $d_{\rm 22\,GHz} <$ 59\,mas (2.8\,pc), appears to be
isolated, and has a very low luminosity, $\nu L_\nu ({\rm 22\,GHz}) =
1.2 \times 10^{35}\,\rm erg\,s^{-1}$, only about 60 times that of
Sagittarius\,A$^{\star}$. Also, the simulated gas flow tailored to
NGC\,3115 developed ADAF traits inside the stagnation radius $R_{\rm
  sta} \sim$ 360\,mas \citep[17\,pc;][]{yao20}.

For the above reasons, we opt to interpret Figure \ref{fig:sed} in
terms of ADAF theory 
\citep[e.g.,][]{NY95a,NY95b,mah97,nem14,ban19,pes21}. In brief, ADAFs
are characterized by two-temperature structures, with their ion
temperatures exceeding their electron temperatures. The electrons are
able to cool via a combination of synchrotron, bremsstrahlung, and
inverse Compton radiation, which together define the SEDs of their
emission observed across the electromagnetic spectrum. Synchrotron 
emission necessitates magnetic fields. These fields are typically
treated as having a tangled geometry and a sub-dominant pressure that
is about a tenth of the gas pressure.

\begin{figure}[tbh]
\plotone{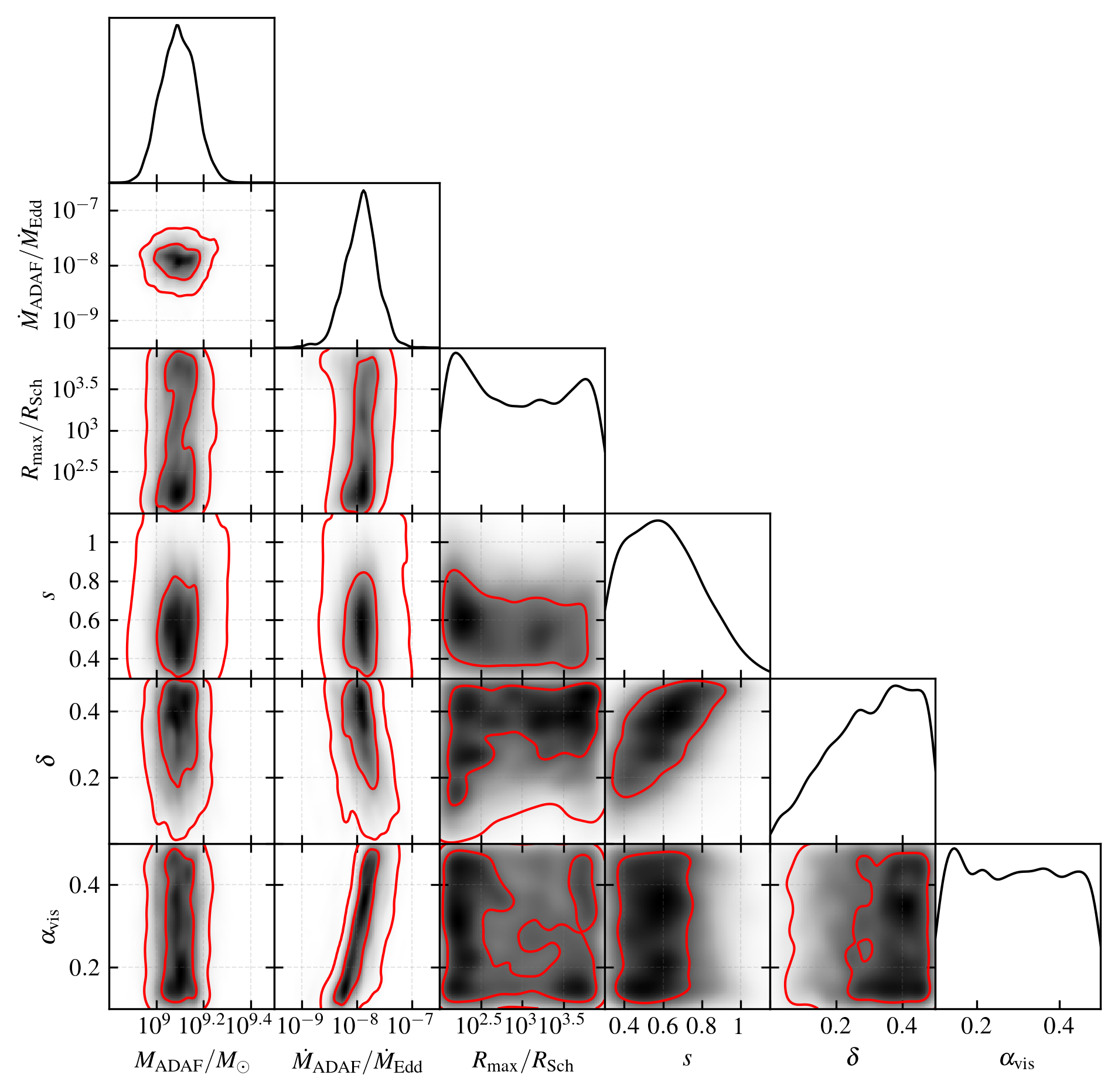}
\caption{Posterior distributions of BH mass $M_{\rm ADAF}$, 
Eddington-scaled accretion rate 
$\dot{M}_{\rm ADAF}/\dot{M}_{\rm Edd}$, 
Schwarzschild-scaled maximum radius $R_{max}/R_{\rm Sch}$, power-law
index for mass accretion rate versus radius $s$, fraction of viscous 
heating going directly to electrons $\delta$, and viscosity parameter 
$\alpha_{\rm vis}$.}
\label{fig:Metc}
\end{figure}

Following \citet{pes21}, we modeled the radio SED peak in Figure
\ref{fig:sed} with a six-parameter fit to a thermal ADAF. (We included
a 3$\sigma$ upper limit of 2.7\,mJy at $\nu =$ 222\,GHz \citep{lo23}
but that datum did not usefully constrain the model.) Table
\ref{tab:Metc} gives the six parameters, their fit priors, and their
fit posteriors, while Figure \ref{fig:Metc} conveys the associated
corner plot. We carried out the fit using the dynesty nested sampling
code \citep{spe20}. The best-constrained parameters for the ADAF are
its BH mass $M_{\rm ADAF}$ and its Eddington-scaled accretion rate 
$\dot{M}_{\rm ADAF}/\dot{M}_{\rm Edd}$ onto the BH. We discuss all the
tabulated parameters in turn below.

\begin{deluxetable*}{clcc}\label{tab:Metc}
\tablecolumns{4} \tablewidth{0pc} 
\tablecaption{ADAF Model for NGC\,3115's SED} 
\tablehead{\colhead{Parameter} & \colhead{Description} &
  \colhead{Range of Prior}     & \colhead{Fit Result}  \\ 
  \colhead{(1)}                & \colhead{(2)}         & 
  \colhead{(3)}                & \colhead{(4)}} 
\startdata
$M_{\rm ADAF}$ & BH mass & $10^8$-$10^{10}$\,$M_{\odot}$ & 
   $(1.2\pm0.2) \times 10^9\,M_\odot$ \\ 
$\dot{M}_{\rm ADAF}/\dot{M}_{\rm Edd}$ & Eddington-scaled 
  accretion rate at  3 $R_{\rm Sch}$ & $10^{-10}$-$10^{-5}$ & 
   $1.2^{+1.0}_{-0.6} \times 10^{-8}$ \\
$R_{\rm max}/R_{\rm Sch}$ & Schwarzschild-scaled maximum radius & 
  $10^2$-$10^4$ & No significant constraint \\
$s$ & Power-law index for mass accretion rate versus radius & 
  0.3-2.0 &  $0.61^{+0.26}_{-0.20}$ \\
$\delta$ & Fraction of viscous heating going directly to electrons & 
  0.01-0.5 &  $0.32^{+0.12}_{-0.15}$ \\
$\alpha_{\rm vis}$ & Viscosity parameter & 
  0.1-0.5 & No significant constraint \\
\enddata
\tablecomments{Three parameters are held fixed: the distance $D =$ 
9.7\,Mpc, the canonical radiative efficiency $\eta = 0.1$, and the
ratio of gas pressure to magnetic pressure $\beta = 10$.}
\end{deluxetable*}

The BH mass $M_{\rm ADAF} = (1.2\pm0.2) \times 10^9\,M_\odot$ inferred
from the ADAF model agrees with previous and independent estimates of
$(1-2) \times 10^9\,M_\odot$ from stellar or hot-gas dynamics
\citep{kor96,ems99,won11,won14}. The Eddington accretion rate and
Schwarzschild radius associated with $M_{\rm ADAF}$ are $\dot{M}_{\rm
  Edd} = 26\,M_\odot$\,yr$^{-1}$ and $R_{\rm Sch} = 1.1 \times
10^{-4}$\,pc, respectively.

For the LLAGN in NGC\,3115, the inferred Eddington-scaled accretion
rate, $\dot{M}_{\rm ADAF}/\dot{M}_{\rm Edd} = 1.2^{+1.0}_{-0.6} \times
10^{-8}$, is about 4-8 times lower than recent estimates from tailored
simulations of Sagittarius\,A$^\star$ \citep{aki22}.

\begin{figure}[tbh]
\plotone{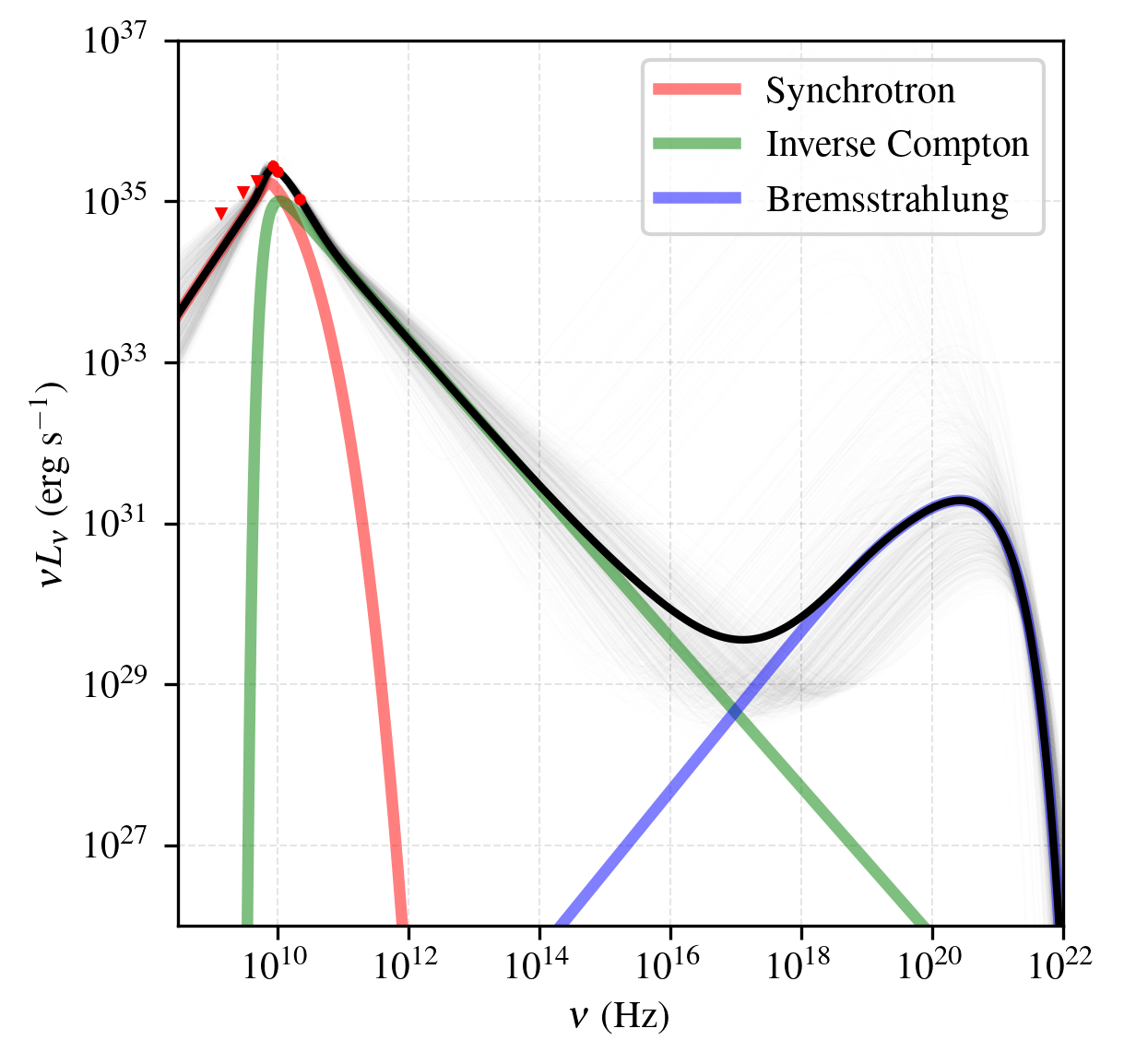}
\caption{$\nu L_\nu$ SED of the LLAGN in NGC\,3115, highlighting
modeled contributions from the individual emission mechanisms across
the electromagnetic spectrum. Figure \ref{fig:sed} is a zoom-in of this
figure's upper-left portion.}
\label{fig:sed_broadband}
\end{figure}

For an ADAF with a very low $\dot{M}_{\rm ADAF}/\dot{M}_{\rm Edd}$, 
Comptonization weakens and the X-ray spectrum can be dominated by
bremsstrahlung emission \citep{yua14}. Our ADAF modeling of the LLAGN
in NGC\,3115 indeed predicts an X-ray bremsstrahlung peak, with a
10-100\,keV luminosity on the order of $10^{31}\,\rm erg\,s^{-1}$ (see
\autoref{fig:sed_broadband}). Unfortunately, the deepest X-ray search
for a LLAGN in NGC\,3115 involves softer wavelengths: $L_{\rm
0.5-6.0\,keV} < 0.44 \times 10^{38}\,\rm erg\,s^{-1}$ and $L_{\rm 
0.5-1.0\,keV} < 0.11 \times 10^{38}\,\rm erg\,s^{-1}$ \citep{won14}.
At harder wavelengths, confusion from NGC\,3115's population of X-ray
binaries led to a limiting value of $L_{\rm 2-10\,keV} < 3.9 \times
10^{38}\,\rm erg\,s^{-1}$ for the LLAGN \citep{wro12}. All of these
upper limits are consistent with our SED fit, which predicts that the
X-ray emission from NGC\,3115's LLAGN should be orders of magnitude
fainter than the sensitivities achieved by the existing observations.

Optical upper limits are available but have luminosities \citep{alm18}
well above the range shown in \autoref{fig:sed_broadband}. Future
high-resolution infrared imaging \citep[e.g.,][]{do14} may help
improve the ADAF fitting.

The peak synchrotron frequency of an ADAF scales roughly as 
$\nu_{\rm peak} \propto M_{\rm ADAF}^{-1/2} \times 
(\dot{M}_{\rm ADAF}/\dot{M}_{\rm Edd})^{1/2}$ \citep{yua14}. For the
LLAGN in NGC\,3115, its extreme pairing of a very high BH mass and a 
very low Eddington-scaled accretion rate places its synchrotron peak
near $\nu_{\rm peak} =$ 9\,GHz, thus at radio wavelengths. This is in
contrast to the oft-studied scenario of ADAFs featuring synchrotron
peaks at millimeter or submillimeter wavelengths 
\citep[e.g.,][]{EHT_MWL21,EHT_MWL22,nai24}.

Other than the BH mass and accretion rate, the remaining parameters in
our SED model are poorly-constrained by the available data (see
\autoref{fig:Metc}). We find a weak constraint on the power-law index
for the mass accretion rate as a function of radius, $s =
0.61^{+0.26}_{-0.20}$, which is consistent with the innermost,
angle-averaged value of $s \sim 0.8$ seen in the gas flow simulation
tailored to NGC\,3115 \citep{yao20}. We find a similarly weak
constraint on the fraction of viscous heating the goes directly to
electrons, $\delta = 0.32^{+0.12}_{-0.15}$. For context, early ADAF
theoretical work \citep[e.g.,][]{NY95a,NY95b,mah97} typically assumed
a value for $\delta$ that is roughly equal to the ratio of the
electron to proton masses, that is, about 1/2000. As the theory has
matured and been tested against observations, considerably larger
values for $\delta$ have become favored, closer to about 0.3
\citep[e.g.,][]{yua14}, which is consistent with our finding. Neither
the maximum Schwarzschild-scaled radius $R_{\rm max}/R_{\rm Sch}$ nor
the viscosity parameter $\alpha_{\rm vis}$ are constrained to any
significant level, and both parameters appear to be entirely
prior-dominated in our fits.

The material forming NGC\,3115's ADAF could affect the linear
polarization of the faint radiation emerging at optically-thin
frequencies $\nu >$ 9\,GHz (\autoref{fig:broadband}). Sensitive
observations with the next-generation VLA \citep[ngVLA;][]{mur18}
could constrain the linear polarization at high resolutions, that is,
with $\theta^\nu_{\rm geo} < 2R_{\rm sta} \sim$ 720\,mas (34\,pc).
Detections or upper limits would guide polarization enhancements to
gas flow simulations or theoretical models. Sufficiently strong
polarization detections would constrain the Faraday rotation measures,
enabling additional tests of simulations or models. Combined with some
assumptions about the accretion flow geometry and magnetic field,
rotation measures would map to accretion rates near the BH
\citep[e.g.,][]{mar06,wie24}.

\section{Beyond the Stagnation Radius}\label{sec:beyond}

Gas motions were predominantly outflows beyond the stagnation radius
$R_{\rm sta} \sim$ 360\,mas (17\,pc) identified in the simulation
tailored to NGC\,3115 \citep{yao20}. From hot-gas modeling, 
\citet{won11} estimated an accretion rate $\dot{M}_{\rm X-ray} = 
2.2 \times 10^{-2}\,M_\odot$\,yr$^{-1}$ at the Bondi radius 
$R_{\rm Bon} = 3\farcs6$ (170\,pc), which is about ten times the 
$R_{\rm sta}$. The accretion rate at the Bondi radius is thus 
already sub-Eddington, $\dot{M}_{\rm X-ray}/\dot{M}_{\rm Edd}
= 10^{-3.1}$. 

This suggests that it could be productive to seek radiative 
signatures, at low resolution ($\theta^\nu_{\rm geo} > 
2R_{\rm sta}$), of material potentially being `lost' from the 
outflow-dominated region. For example, deep radio continuum images
could trace off-nuclear components. No off-nuclear emission has been
reported from the low-resolution images in Table \ref{tab:augment}, 
but the RMS values of those images are quite modest 
\citep{fab89,whi97,lac20}. Additionally, X-ray emission lines could
trace off-nuclear hot winds resembling those recently reported for
M81 and NGC\,7213 \citep[e.g.,][]{shi21,shi22}.

\section{Summary and Conclusions}\label{sec:summary}

NGC\,3115 hosts the nearest billion-solar-mass BH, with a fiducial 
mass $M_{\rm fid} \sim 1.5 \times 10^9\,M_\odot$ from stellar or 
hot-gas dynamics \citep{kor96,ems99,won11,won14}. A hydrodynamic 
simulation tailored to NGC\,3115 found that gas motions were
predominantly inflows inside the stagnation radius $R_{\rm sta} 
\sim$ 360\,mas (17\,pc) and developed ADAF-like characteristics
\citep{yao20}. We observed the LLAGN in NGC\,3115 with the JVLA at
a high resolution, taken to mean $\theta^\nu_{\rm geo} < 
2R_{\rm sta}$. This approach mitigated contamination from emission
from the outflow-dominated zone beyond $R_{\rm sta}$. Our principal
findings are:

\begin{enumerate}

\item For NGC\,3115, we localized a faint continuum nucleus, with flux
  density $S_{\rm 22\,GHz} = 48.2\pm6.4\,\mu$Jy, to a FWHM diameter
  $d_{\rm 22\,GHz} <$ 59\,mas (2.8\,pc). We found no evidence for
  adjacent emission within the stagnation radius $R_{\rm sta}$. The
  nucleus' luminosity density $L_{\rm 22\,GHz} = 5.4 \times
  10^{17}\,\rm W\,Hz^{-1}$ is about 60 times that of
  Sagittarius\,A$^\star$.

\item We augmented the new $S_{\rm 22\,GHz}$ with previous continuum
  photometry for NGC\,3115 at lower radio frequencies.  A detection at
  high resolution implied the nucleus has a spectral index
  $\alpha_{\rm 10\,GHz}^{\rm 22\,GHz} = -1.85\pm0.18$ ($S_\nu \propto
  \nu^\alpha$), indicating optically-thin synchrotron
  emission. Folding in upper limits at low resolution $\theta^\nu_{\rm
    geo} < 2R_{\rm sta}$, the SED of the nucleus peaks near $\nu_{\rm
    peak} =$ 9\,GHz.

\item For NGC\,3115, we modeled its SED as an ADAF and inferred a BH
  mass $M_{\rm ADAF} = (1.2\pm0.2) \times 10^9\,M_\odot$, consistent
  with independent fiducial estimates. We also inferred an
  Eddington-scaled accretion rate, $\dot{M}_{\rm ADAF}/\dot{M}_{\rm
    Edd} = 1.2^{+1.0}_{-0.6} \times 10^{-8}$, which is about 4-8 times
  lower than recent estimates for Sagittarius\,A$^\star$.

\end{enumerate}

An important next step for NGC\,3115 is to conduct deep polarimetric
imaging using radio facilities. This will improve the SED of the 
LLAGN, enable a search for very faint outflows, and constrain Faraday
rotation measures to guide magnetic-field enhancements to the 
tailored gas-flow simulations. In addition, other absorption-line 
nuclei in optically bright galaxies should be searched for analogs of
the LLAGN in NGC\,3115. If their BHs are less massive than NGC\,3115's,
we speculate that their ADAFs, if present, could exhibit SEDs that peak
at frequencies $\nu_{\rm peak} >$ 9\,GHz.

\begin{acknowledgments}
We thank the anonymous reviewer for their helpful and timely
comments. The NRAO is a facility of the National Science Foundation
(NSF), operated under cooperative agreement by Associated
Universities, Inc. (AUI). The ngVLA is a design and development
project of the NSF operated under cooperative agreement by AUI. The
new JVLA data used in this study may be obtained from the NRAO Data
Archive (https://data.nrao.edu) via project code 24B-030 (PI
J. Wrobel).

Support for D.W.P. was provided by the NSF through grants AST-1935980 and AST-2034306, and by the Gordon and Betty Moore Foundation through grants GBMF5278 and GBMF10423.  The Black Hole Initiative at Harvard University is funded by the John Templeton Foundation (grants 60477, 61479, and 62286) and the Gordon and Betty Moore Foundation (grant GBMF8273). 
Basic research in radio astronomy at the U.S. Naval Research Laboratory is supported by 6.1 Base Funding.
\end{acknowledgments}

\begin{contribution}
Author order reflects contribution effort.
\end{contribution}

\facilities{CXO - Chandra X-ray Observatory satellite, VLA - Very Large Array}

\software{CARTA (Comrie et al. 2021), CASA (The CASA Team 2022), dynesty \citep{spe20},
matplotlib \citep{matplotlib}, numpy \citep{numpy_2011,numpy_2020}}

\appendix
\numberwithin{equation}{section}

\section{Source size modeling} \label{app:SourceSizeModeling}

To determine the angular size of the emitting region reported in \autoref{sec:data}, we fit a circularly symmetric Gaussian source structure model to the calibrated visibility data.  Our model for the visibility structure $V(u,v)$ as a function of frequency $\nu$ is given by

\begin{equation}
V(u,v,\nu) = S \left( \frac{\nu}{\text{22\,GHz}} \right)^{\alpha} e^{- 2 \pi i (ux_0 + vy_0)} \exp\left[ -2 \pi^2 \sigma_G^2 (u^2 + v^2) \right] , \label{eqn:GaussianModel}
\end{equation}

\noindent where $S$ and $\alpha$ are the flux density and spectral index of the source measured at 22\,GHz, $(x_0,y_0)$ is its coordinate location relative to the phase center of the observations, and $\sigma_G$ is the Gaussian width (i.e., angular extent) of the source on the sky.  These model parameters are related to the quantities reported in \autoref{sec:data} by

\begin{subequations}
\begin{align}
S_{\rm 22\,GHz} & = S \\
\alpha_{\rm 18\,GHz}^{\rm 26\,GHz} & = \alpha \\
d_{\rm 22\,GHz} & = 2 \sqrt{2 \ln(2)} \sigma_G .
\end{align}
\end{subequations}

Prior to fitting, we average the calibrated data in time on 10-minute intervals and in frequency across spectral windows (64 in total, 128\,MHz each).  For each visibility data point $\hat{V}$, our likelihood function $\ell_k$ is a complex Gaussian function,

\begin{equation}
\ell_k = \frac{1}{2 \pi \sigma_k^2} \exp\left( - \frac{1}{2 \sigma_k^2} \left| V(u_k,v_k,\nu_k) - \hat{V}(u_k,v_k,\nu_k) \right|^2 \right) ,
\end{equation}

\noindent where $\sigma_k$ is the uncertainty in the measurement and $V(u_k,v_k,\nu_k)$ is given by \autoref{eqn:GaussianModel}.  We assume that the measurement uncertainty $\sigma_k$ is the same for all data points, and we include this quantity as an additional parameter in our model.  The total likelihood function $\mathcal{L}$ over all data points is then the product of the individual likelihoods,

\begin{equation}
\mathcal{L} = \prod_k \ell_k .
\end{equation}

We specify uniform priors of $x_0 \sim [-0.05,0.05]$ and $y_0 \sim [-0.05,0.05]$ (in units of arcseconds), $S \sim [0,100]$ (in units of mJy), $\alpha \sim [-5,0]$, $\sigma_G \sim [0,0.05]$ (in units of arcseconds), and $\sigma_k \sim [1,5]$ (in units of mJy).  We used the nested sampling package dynesty \citep{spe20} to carry out the parameter space exploration, for which results are summarized in \autoref{sec:data} and shown in \autoref{fig:Setc}.

\bibliography{ngc3115ad}{}
\bibliographystyle{aasjournal}

\end{document}